\def\year{2021}\relax
\documentclass[letterpaper]{article} 
\usepackage{aaai21}  
\usepackage{times}  
\usepackage{helvet} 
\usepackage{courier}  
\usepackage[hyphens]{url}  
\usepackage{graphicx} 
\usepackage{natbib}  
\usepackage{caption} 
\usepackage{color}

\usepackage{bbm}
\usepackage{amsmath}
\usepackage{lipsum}
\usepackage{enumitem}
\usepackage{placeins}

\title{Emotion Regulation and Dynamics of Moral Concerns During the Early COVID-19 Pandemic}
\author{
    Siyi Guo\textsuperscript{\rm 1}\textsuperscript{\rm 2},  Keith Burghardt\textsuperscript{\rm 2}, Ashwin Rao\textsuperscript{\rm 1}\textsuperscript{\rm 2}, and Kristina Lerman\textsuperscript{\rm 2}\\
}
\affiliations{
    \textsuperscript{\rm 1}Department of Computer Science, University of Southern California\\
    \textsuperscript{\rm 2}Information Sciences Institute, University of Southern California\\

    \{siyiguo,keithab,ashreyas,lerman\}@isi.edu\\
}

\begin{document}

\maketitle

\begin{abstract}
The COVID-19 pandemic has upended daily life around the globe, posing a threat to public health. Intuitively, we expect that surging cases and deaths would lead to fear, distress and other negative emotions. However, using state-of-the-art methods to measure sentiment, emotions, and moral concerns in social media messages posted in the early stage of the pandemic, we see a counter-intuitive rise in positive affect. 
We hypothesize that the increase of positivity 
is associated with a decrease of uncertainty and emotion regulation. Finally, we identify a partisan divide in moral and emotional reactions that emerged after the first US death. 
Overall, these results show how collective emotional states have changed since the pandemic began, and how social media can provide a useful tool to understand, and even regulate, diverse patterns underlying human affect.
\end{abstract}

\section{Introduction}
Emotions shape beliefs~\cite{ostrom1969relationship} and moral reactions to social dilemmas~\cite{haidt2007moral,10.1371/journal.pone.0253326}, 
promote group identity \cite{graham2009liberals}, especially in time of crisis, and help people understand appropriate social response to conflict~\cite{vanKleef2016Editorial}. 
As social interactions have moved online en masse, the transition creates new opportunities to study belief dynamics at an unprecedented spatial scale and temporal resolution. 
Despite some well-known limitations of social media data~\cite{Ruths2014}, it can help researchers study the interplay between emotions, beliefs and attitudes, and supplement traditional instruments to survey public opinion~\cite{Lwin2020Global,li2020impact}. We aim to understand the emotional reactions to the COVID-19 pandemic on social media. The pandemic brought about enormous societal disruptions, such as lockdowns, school closures and transitions to online learning, which had a profound effect on some emotions people expressed~\cite{Lwin2020Global}. 
To understand the population's collective emotional states, needs, concerns and psychological well-being during the pandemic, researchers have used 
data-driven approaches, applying natural language processing (NLP) methods to social media data 
(e.g., \citealt{Lwin2020Global,aiello2021epidemic,ashokkumar2021social}). 


In this work, we measure dynamics of emotions and moral concerns during the COVID pandemic. We analyze English-language messages about the pandemic on Twitter between January and May 2020. We use state-of-the-art language models to measure emotions, including fear, anger, and optimism, and assess intensity of moral reactions along  dimensions such as care, harm, fairness and cheating. We compare results to simpler dictionary-based model that quantify emotions and sentiment. 
While each model has limitations, such as data shift that can affect emotion or moral label accuracy \cite{Elsahar2019-annotate}, the consistency of results suggests a robust effect. 
We also split heterogeneous population of users along the ideological line and separately study each user group's emotional and moral dynamics. We find diverging reactions due to both the pandemic and the ideological differences. 

Our main findings are the following.
\begin{enumerate}
    \item Positive moral concerns (virtues) and positive emotions increased while fear decreased during early stage of the pandemic, despite that disgust, anger and fear were the dominant emotions.
    \item The counter-intuitive results were found to be robust to different text-classification models.
    \item The changes in emotion were correlated with a decrease in uncertainty and increase in solidarity (loyalty \& care).
    \item Political differences mediate emotional and moral reactions, and we identify a partisan divide in affect that emerged after the first US death.
\end{enumerate}
The trends in positive emotions suggest that people used social media to regulate negative emotions. Meanwhile, we find a reduction in uncertainty about the pandemic might also decrease fear and increase positive emotions. Finally, the pandemic has enhanced the emotional divide between ideologies, which may explain increasing ideological polarization.  
These results bring new understanding to how emotions are regulated after major global crises and illuminate the role emotions play in shaping attitudes. 

\vspace{-5pt}
\section{Related Works}


Previous studies have explored the psychological impact of COVID-19 pandemic, using surveys to show an increase in distress and uncertainty~\cite{stress_in_america}. Other works focused on data-driven methods, utilizing search data or Wikipedia data \cite{suh2021population,ribeiro2021sudden}. 
Researchers have also analyzed social media data to study people's sentiment, emotions and mental health concerns. Lwin et al. (\citeyear{Lwin2020Global}) found that public emotions in Twitter shifted from fear to anger  early in the pandemic, along with increasing sadness and joy. Other researchers found that negative emotions increased and positive emotions and life satisfaction decreased among Chinese social media users \cite{li2020impact}. 
Biester et al. (\citeyear{10.1145/3458770}) and Saha et al. (\citeyear{info:doi/10.2196/22600}) focused on studying the impact of COVID-19 on people's mental health concerns and diseases, such as depression and suicidal ideation. Aiello et al. \citeyear{aiello2021epidemic} and Ashokkumar et al. \citeyear{ashokkumar2021social} each studied how people's mental states shifted through different phases since the pandemic began, and similarly found people's transition from refusal or warning, anger to acceptance of the ``new-normal.'' 

With respect to measuring moral concerns, the Moral Foundation Theory is widely adopted \cite{graham2009liberals}. It defines five moral foundations: \textit{care/harm} (empathy, dislike of suffering), \textit{fairness/cheating} (proportionality, justice and rights), \textit{loyalty/betrayal} (sense of belonging to identified group), \textit{authority/subversion} (respect of authority and tradition), and \textit{sanctity/degradation} (concerns with sickness and contamination). Previous work on moral foundations and COVID-19 includes Ekici et al. (\citeyear{ekici2021deciding}), who showed that fairness, care, and purity moral foundations are found to be the most relevant to COVID-19. Henderson et al. (\citeyear{henderson2021disease}), in contrast, found that individuals who were worried about disease infection made harsher moral judgments.
Finally, prior works have discussed the heterogeneity in social media, especially between political ideologies. 
Researchers have found divergent language use on social media during the COVID-19 pandemic between the partisan ideologies \cite{gollwitzer2020partisan}. Discussions in the US, for example about government measures, is largely shaped by political polarization \cite{rao2021political,Schmitz2022}. In addition, conservative elites have consistently downplayed the dangers of COVID-19 \cite{10.1371/journal.pone.0253326}. 






\vspace{-5pt}
\section{Methods}
\vspace{-4pt}
\subsection{Data}

We use an open source COVID-19 tweet data \cite{info:doi/10.2196/19273} collected based on a list of COVID-19 related keywords such as ``coronavirus,'' ``pandemic,''  and ``stay-at-home.'' We select all English tweets posted between 01/24 to 05/01/2020 from the US within this corpus. The refined dataset consists of 27M tweets posted by 2.4M users.


\vspace{-4pt}
\subsection{Quantifying Emotions and Moral Concerns}
To quantify emotions expressed in tweets, we use SpanEmo \cite{alhuzali-ananiadou-2021-spanemo}, a state-of-the-art performing model on a benchmark dataset SemEval 2018 Task 1e-c \cite{mohammad-etal-2018-semeval}. This BERT-based model accounts for correlations among the emotions to improve performance. The emotions we measure are \textit{anger},  \textit{disgust}, \textit{fear}, \textit{pessimism}, \textit{sadness}, \textit{joy}, \textit{love}, \textit{anticipation}, \textit{optimism}, \textit{surprise} and \textit{trust}. One tweet could be labeled with multiple emotions. We calculate the fractions of tweets labeled with each emotion in each day to construct the time series.

To measure moral concerns, we follow a prior work \cite{10.1007/978-3-030-60975-7_16} and use a dictionary-based approach. We use the extended moral foundation dictionary \cite{Hopp_Fisher_Cornell_Huskey_Weber_2020} to define the words associated with the virtues and vices (e.g. care and harm) and use an embedding-based method, FrameAxis~\cite{kwak2021frameaxis}, to measure the moral \textit{bias} (whether the text tends to discuss the vice or virtue of a morality) and \textit{intensity} (whether the text has a greater or lesser focus on that moral foundation). With the numerical scores of bias and intensity, we take the daily averages from all tweets to construct the time series, and then analyse the collective expressions of emotions and morals. 

\vspace{-4pt}
\subsection{Quantifying Uncertainty}

To measure the degree of uncertainty in tweets, we use the Uncertainty Corpus \cite{durieux_2019}, which contains terms such as ``probably'', ``confused'', ``not sure''. For each date, we perform a keyword search for these terms in all English tweets and calculate the fraction of tweets that contain uncertainty terms out of all tweets. To reduce computation time complexity, we measure uncertainty in 5\% uniformly random sub-sample of all tweets in our dataset.

\vspace{-4pt}
\subsection{Quantifying Partisan Differences}

Political polarization shapes many aspects of the US society, including discussions and reactions to the COVID-19 pandemic. In order to explore this heterogeneity in the data, we quantify the political ideology of user groups using a method recently used in previous work \cite{rao2021political}. We extract media URLs posted in tweets, whose political leanings were classified by neutral websites, to infer the political ideology (liberal and conservative) of users. Next, we train a logistic regression model using fastText embeddings of tweets from these labeled users. For other users who's posts don't contain enough informative URLs, we use this classifier to infer their ideologies. We then aggregate all tweets posted by different user groups on each day and analyze their collective emotions and moral foundations.

\vspace{-5pt}
\section{Results}

\begin{figure*}[h]
\centering
\includegraphics[width=0.85\textwidth]{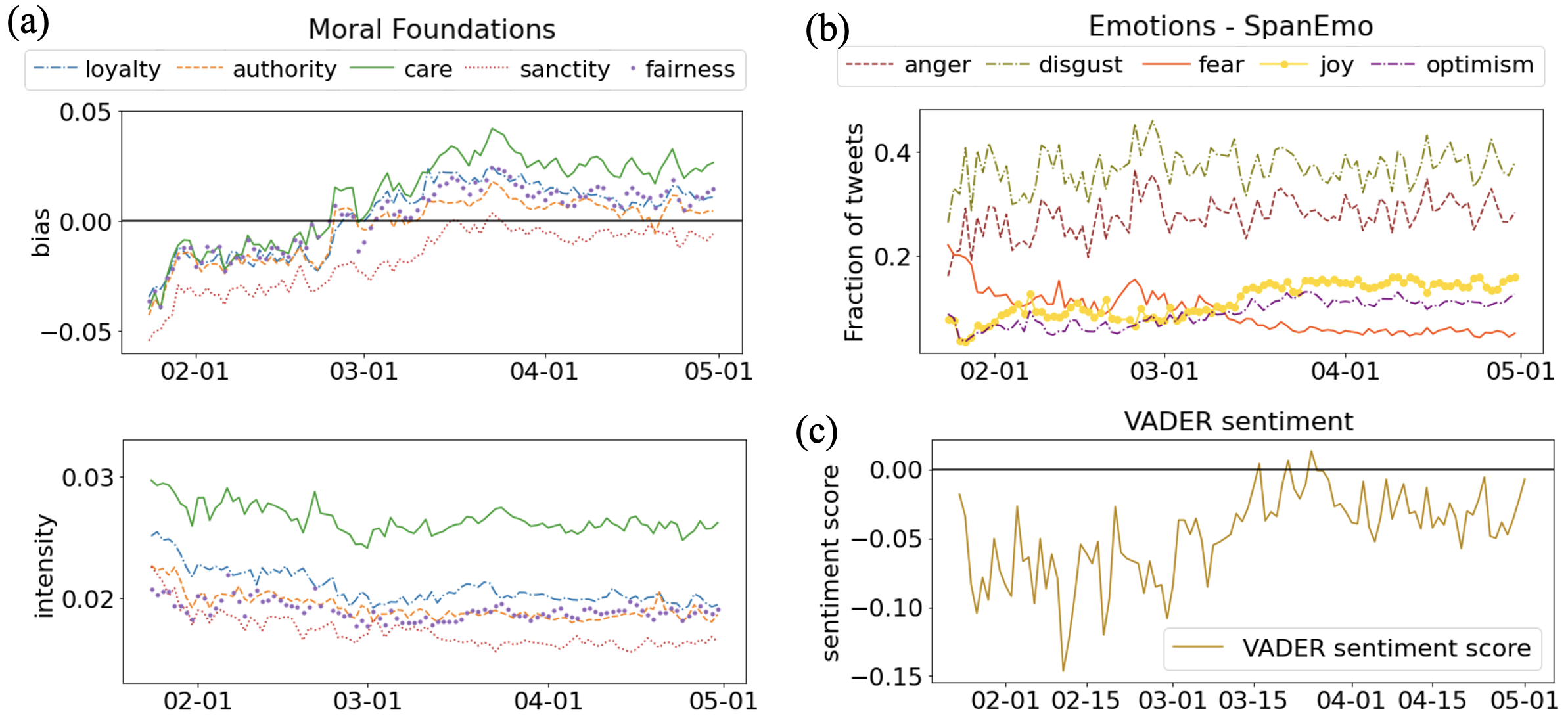}
\caption{The trend of people dominantly expressing negativity but also growing positive feelings is shown with time series of (a) daily-averaged bias and intensity scores of moral foundations, (b) daily fraction of tweets labeled with different emotions by SpanEmo from 01/24/2020 to 05/01/2020. Other measured emotions not shown are anticipation and sadness, which remain steady of the time frame, and love, pessimism, surprise and trust, each of which is less than 1\% of the daily tweets. (c) The daily averaged VADER compound sentiment scores. It reflects a similar trend as of moral foundations and emotions.}
\label{fig_time_series}
\end{figure*}
\vspace{-4pt}
\subsection{Dynamics of Emotions and Moral Concerns}

Figure~\ref{fig_time_series}a, b show the time series of the daily averaged bias and intensity of moral foundations, and the fraction of daily tweets containing each emotion during this period. Analysis of five foundations of moral concerns in Fig.~\ref{fig_time_series}a shows that \textit{care/harm} foundation had the strongest expression, i.e., highest intensity, followed by \textit{loyalty/betrayal} (Fig.~\ref{fig_time_series}a). The higher intensity of the \textit{care/harm} foundation, coupled with the negative bias, implies expressions of negative sentiment (or vices) were more common. This suggests that users focused on harms through the end of February. Similarly, the high intensity of the \textit{loyalty}/\textit{betrayal} foundation, coupled with negative bias, suggests that people were expressing concerns about outsiders. As for the \textit{sanctity}/\textit{degradation} foundation related to concerns about contamination and infection with the virus, the bias remained mostly negative. We also find the bias of the remaining moral foundations increased, meaning that people expressed more moral virtues from mid-February to the end of March. Finally, the intensity of all moral foundations decreased, suggesting the frequency of moral discussions decreased. 

Fig.~\ref{fig_time_series}b shows the time series of selected emotions. Out of ten emotions we measured, \textit{disgust} and \textit{anger} are the most dominant emotions, expressed in about a third of all tweets. People hold strong negative emotions about the virus and the unfolding pandemic, and these results also reflect a tendency to dwell on negative emotions, known as the negativity bias \cite{fessler2014negatively}. Interestingly, we see that \textit{fear} decreased over March 2020 while some of the positive emotions (\textit{optimism} and \textit{joy}) increased during this time period. This can be counter-intuitive, as we might expect people getting more fearful and negative as the number of cases rising and US saw its first COVID-19 death. Meanwhile, we also observe the rise of the virtues of moral foundations, especially \textit{loyalty} and \textit{care}, suggesting that rise in positive emotions co-occurred with rising solidarity.


\paragraph{Validity of emotion and moral measurements}

\begin{figure}[!h]
    \centering
    \includegraphics[width=0.9\linewidth]{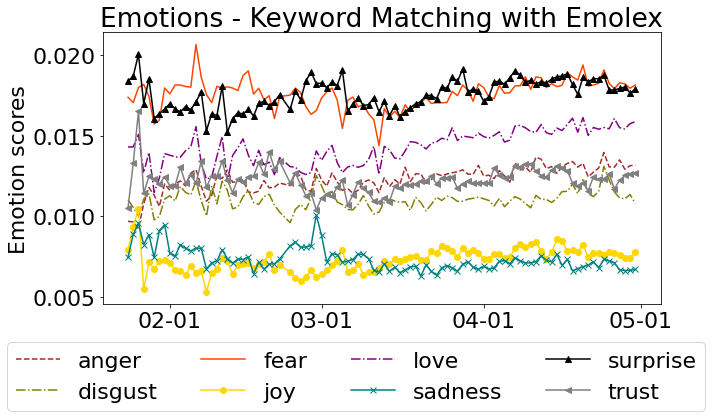}
    \caption{Trend in emotions extracted by keyword matching using EmoLex. Daily average is taken in the time series.}
    \label{fig_emolex}
\end{figure}

Manual annotation is the gold standard for evaluating the performance of models, such as SpanEmo; however, it is often not feasible. Instead, we explore validity of results by comparing them to other methods on 5\% of uniformly sampled data and looking at internal consistency. We use dictionary-based method with EmoLex \cite{Mohammad13}, a widely-used lexicon of emotion-laden words (see \citealt{aiello2021epidemic,Wang2021EmotionAV}) to measure eight emotions (Fig.~\ref{fig_emolex}). The score for each emotion is the number of emotion words appearing in a tweet, normalized by the total number of words in the tweet. The results suggest that \textit{fear} and \textit{surprise} are the top two emotions, in contrast to top emotions in SpanEmo results, \textit{anger} and \textit{disgust}. Still, \textit{fear} is the third most common emotion according to SpanEmo, and concordance of these results is consistent with negativity bias \cite{fessler2014negatively}. 
Further, in both SpanEmo and EmoLex, similar trends are observed, such as increasing \textit{joy} and decreasing \textit{sadness} after the first US death.

We also perform sentiment analysis using a lexicon-based method called VADER, that is specialized to social media text \cite{hutto2014vader}. The compound score sentiment is a measure of positive or negative affect. The time series conveys a similar message as the trends we observe in emotions and moral reactions (Fig.~\ref{fig_time_series}c). The overall sentiment is negative but with an increasing trend especially from 02/15 to the end of March (03/24/2020). To check whether the increase in positive sentiment was due to a change in user composition, we run the same analysis on a group of users who were active before 02/15/2020, and track their moral foundations and emotions from 01/24/2020 to 05/01/2020. We see consistent trends of increasing virtues and positive emotions, and decreasing moral intensity and fear.
The overall consistency between emotions and moral reactions measured by diverse methods give us confidence about the validity of the measurements.

\vspace{-4pt}
\subsection{Uncertainty and Fear of the Unknown}

\begin{figure}
    \centering
    \includegraphics[width=0.45\textwidth]{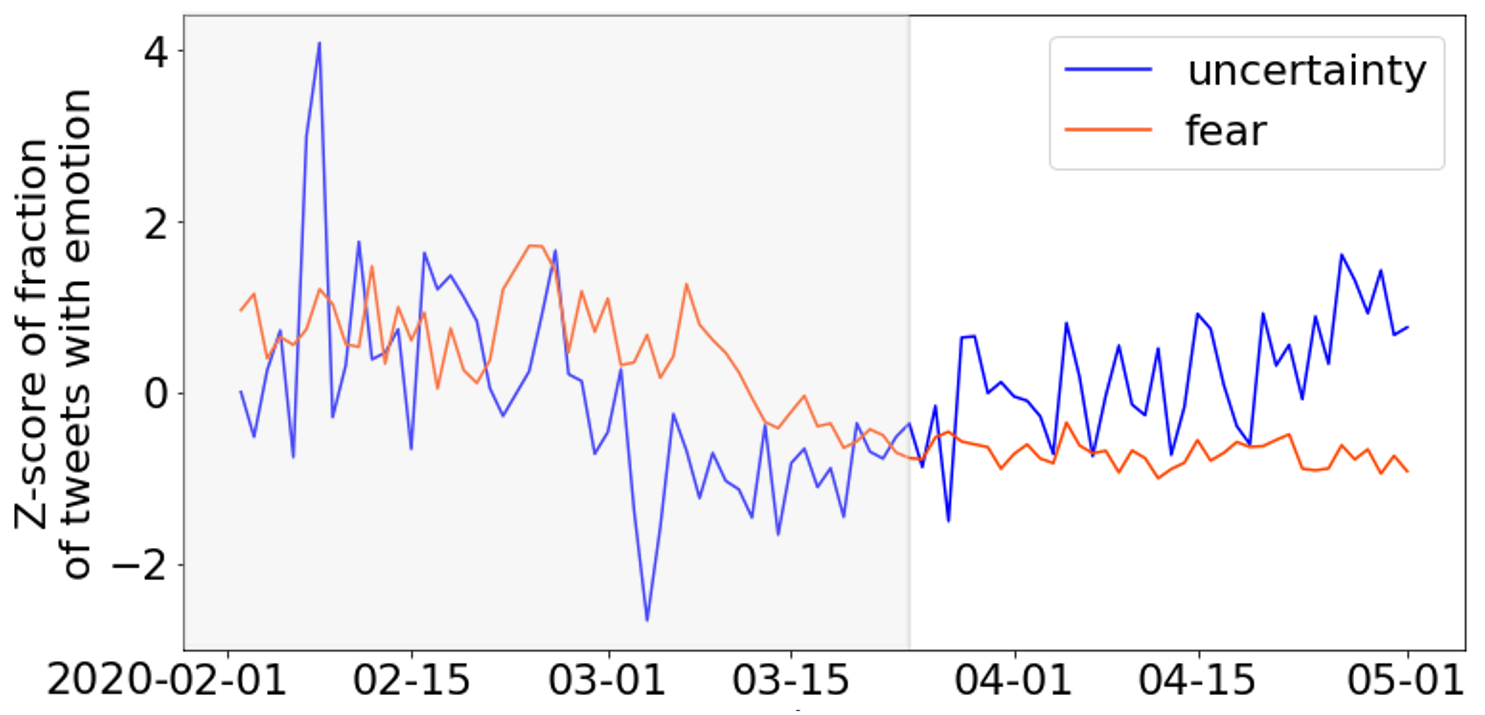}
    \caption{The \textit{Z score} of fraction of daily tweets with fear is associated with that of the uncertainty emotions in 5\% randomly sampled data subset. The fear emotion and the uncertainty emotion are the most correlated (Spearman correlation $= 0.42$, $p$-value $= 0.002$) during the shaded time period.}
    \label{fig_uncertainty}
\end{figure}

We found that the number of tweets expressing \textit{fear} decreased by the end of March 2020, which could be due to the fear of the unknown \cite{CARLETON20165}. As the pandemic started, there were many unknowns about the disease, such as transmission mechanisms, how deadly it is, treatments and the preventive strategies. We hypothesize that people might be more afraid early in the pandemic because of these uncertainty. As epidemiologists and doctors learned more about the disease, the fear of the unknown subsided. 
To verify this hypothesis, we measure the feelings of unknown or uncertainty people expressed in tweets (see Methods). We found that \textit{fear} decreased substantially before the lockdowns happened around 03/24/2020, and the number of tweets expressing \textit{uncertainty} also decreased (shaded area in Figure \ref{fig_uncertainty}). Spearman correlation between fear and uncertainty of the shaded region is $0.42$, $p$-value $= 0.002$ (the same correlation is $0.38$ using EmoLex method to measure fear, $p$-value $= 0.003$). We also find that among all tweets labeled with fear, 26\% also express uncertainty. Therefore the March 2020 decrease in fear is correlated with less uncertainty. Uncertainty or unpredictability can bring higher anxiety and stress \cite{10.3389/fnint.2011.00055}, 
therefore the decrease uncertainty can lead to increasing positive emotions we observe. We see that after the 03/24/2020 lockdowns started, \textit{uncertainty} and \textit{fear} decoupled. The fear stayed low and did not change much, whereas the uncertainty started to increase again, possibly because people were unsure of the impact of the lockdowns, but were emotionally adapting to them. 

\vspace{-4pt}
\subsection{Partisan Differences in Emotions and Morals} 

\begin{figure*}[t]
\centering
\includegraphics[width=1\textwidth]{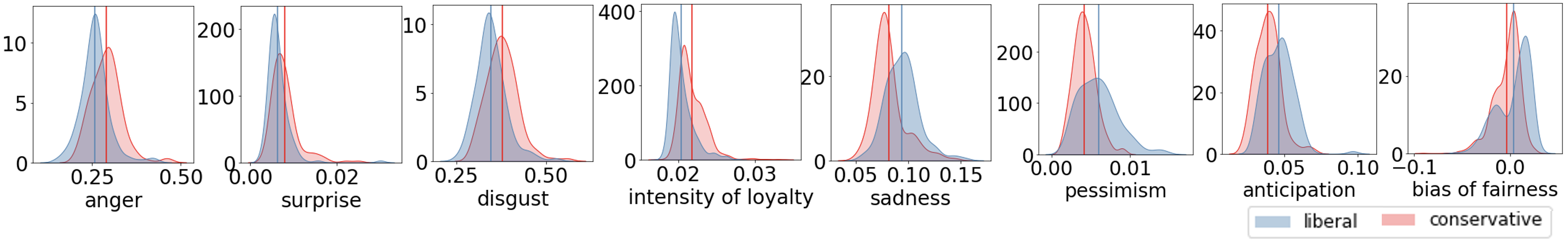}
\caption{The heterogeneity of emotional and moral reactions due to ideological differences. Each density plot consists of all measurements in the labeled variable and labeled population from 01/24/2020 to 05/01/2020. The vertical blue and red lines are the distribution means.}
\label{fig_lib_conserv}
\end{figure*}

\begin{figure}[h!]
\centering
\includegraphics[width=0.45\textwidth]{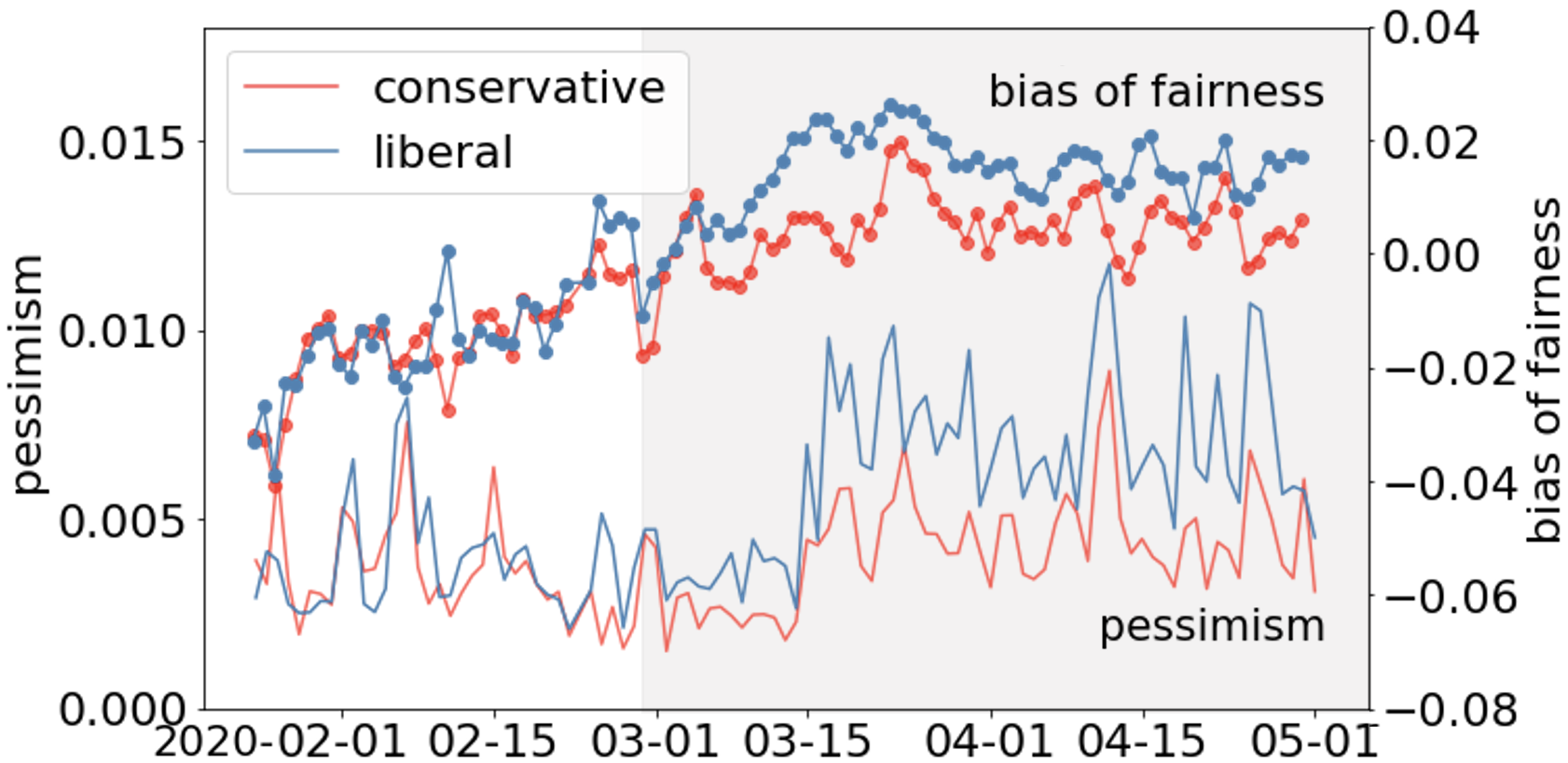}
\caption{Examples of how emotions and moral foundations in different populations evolve. Bias of fairness foundation and the pessimism emotion both started to diverge in liberals and conservatives after the first death happened on 02/28/2020, indicated by the shaded area.}
\label{fig_lib_conserv_ts}
\end{figure}

Social media is highly heterogeneous, with users with different ideologies and concerns. 
Prior research identified strong ideological differences in the attitudes toward the pandemic: liberals and conservatives have different perceptions of pandemic's severity and efficacy of public health interventions~\cite{gollwitzer2020partisan,10.1371/journal.pone.0253326}. 

We split users in our sample according to their partisan ideology (see Methods) and identify differences in their emotional and moral expressions. Figure~\ref{fig_lib_conserv}a shows emotions and moral foundations with significantly different distributions between liberals and conservatives (T-test p-values $<0.01$ between respective distributions). We find that conservative users expressed on average more \textit{surprise}, \textit{anger}, \textit{disgust} and a higher intensity for the \textit{loyalty} foundation compared to liberals. On the other hand, we find that liberals were more \textit{sad} and \textit{pessimistic}, but also had higher \textit{anticipation} and higher bias of \textit{fairness} than conservatives. These observations agree with Jost et al. (\citeyear{jost2003political}): conservatism arises from the need to manage anxiety and threat with its concomitant emotions such as anger and disgust, therefore conservatives have higher \textit{disgust} sensitivity than liberals \cite{terrizzi2013behavioral} and put a greater emphasis on the binding moral foundations including \textit{loyalty} \cite{VANLEEUWEN2009169}. On the other hand, liberal thinking is associated to more suspicion, pessimism and cynicism \cite{doi:10.1177/1369148119866086}, and liberals have greater endorsement and use of the \textit{fairness} foundations \cite{graham2009liberals}. 

Figure~\ref{fig_lib_conserv_ts} shows that \textit{pessimism} and bias of \textit{fairness} behaved similarly between ideologies until the end of February 2020, when these metrics for liberals and conservatives started to diverge  (shaded area). We also find this phenomenon in some other emotions and moral foundations that differ significantly between ideologies (T-test p-values $<0.05$), such as anger, sadness and bias of care foundation. This could be because there was little partisan divide when making sense of the situation, but different subgroups quickly began to express moral foundations and emotions in line with their different identities and ideological attitudes, when critical events occurred.

\vspace{-5pt}
\section{Discussion}
The seemly counter-intuitive trends we have observed, such as the decrease of fear and uncertainty and the increase of positivity, could be related to several psychological theories, such as emotion regulation and uncertainty reduction. Previous research has found that when an event happens, people would shift their positive or negative perceptions to be more neutral, a phenomenon known as hedonic adaptation \cite{kahneman_diener_schwarz_2003}. 
We hypothesize that when facing a terrifying new pandemic, negative emotions and moral concerns decreased over time as people changed their perceptions to be more neutral. The decreasing intensity of moral discussions (bottom fig. \ref{fig_time_series}a) also suggests users adapting to the situation. Another related explanation for the decrease in negative emotion is emotion regulation, i.e., up-regulating the positive emotions and decreasing the negative emotions \cite{BLUMBERG2016105}, due to social media's ability to exchange social support, and increase bonding in lonely times \cite{oh2014does}. 
Our observations on emotions and moral concerns also reflect how people's mental states shifted through different phases since the pandemic began. Aiello et al. (\citeyear{aiello2021epidemic}) showed that people went through (1) the refusal phase, in which people were warned but refused to accept COVID happening, (2) the anger phase, in which fear decreased after the first US death and transitioned to anger, and (3) the acceptance phase beginning after the lockdowns, in which people started to accept the ``new normal.'' We similarly observe the trends of decreasing fear, decreasing moral discussion intensity and increasing positive affect, which are strikingly consistent with these phase transitions.

\vspace{-5pt}
\section{Conclusions}
Our results highlight the utility of social media as a sensor of collective affect. Applying the methods to automatically recognize  emotions and moral concerns in online discussions about the COVID-19 pandemic, we observe a rise in positive affect and a decrease in fear early in the pandemic. 
We associate these trends with the reduction in uncertainty about the pandemic through collective sense-making as well as newfound solidarity. Another key observation is how the pandemic has amplified ideological polarization. We observe ideological differences in emotional and moral reactions, 
and see that emotional expressions of conservatives began to diverge from those of liberals after the first US COVID-19 death. Understanding emotions and morals will help us better explain the dynamics of attitudes and polarization. 

There were, however, a number of limitations of this study. First, we have no ground truth emotions, and must rely on dictionaries and models to measure moral foundations and emotions at scale. 
In addition, the method we use to determine political ideology can make mistakes \cite{rao2021political}, which could affect the differences we see in various groups. Finally, our results were gathered from tweets containing particular COVID-19 keywords but it is possible the keywords do not fully capture the users discussing COVID-19, nor can we be sure that the results are representative to users outside of this cohort, or outside of social media.
The limitations demonstrate a need to extend these results to other more representative datasets and to determine the robustness of these results with findings from human annotators. 

\vspace{-4pt}
\subsection{Ethical Considerations}
The study was reviewed by an institutional review board and found to be exempt. To preserve anonymity and to follow Twitter's terms of service, the data was studied at an aggregate level, therefore we do not anticipate any negative ethical impact from these results. Instead, these results can give insights into emotional well-being of users after pandemics strike. Diverging differences in viewpoints between users in social media can also be taken into account in future public health policies.

\vspace{-5pt}
\section*{Acknowledgements}
This work was funded, in part, by DARPA under contract HR001121C0168.

\vspace{-5pt}
\small


\end{document}